\documentstyle[11pt]{article}
\begin{document}
\begin{large}
\begin{titlepage}

\vspace{0.2cm}

\title{Pair production of neutralinos via gluon-gluon collisions
\footnote{The project supported by National Natural Science
          Foundation of China}}
\author{{Jiang Yi$^{b}$, Ma Wen-Gan$^{a,b}$, Han Liang$^{b}$, Yu Zeng-Hui$^{c}$
        and H. Pietschmann$^{c}$ }\\
{\small $^{a}$CCAST (World Laboratory), P.O.Box 8730, Beijing 100080,
China.}\\
{\small $^{b}$Department of Modern Physics, University of Science
        and Technology}\\
{\small of China (USTC), Hefei, Anhui 230027, China.}\\
{\small $^{c}$ Institut f\"{u}r Theoretische Physik, Universit\"{a}t Wien,
A-1090 Vienna, Austria} }
\date{}
\maketitle

\vskip 12mm

\begin{center}\begin{minipage}{5in}

\vskip 5mm
\begin{center} {\bf Abstract}\end{center}
\baselineskip 0.3in
{The production of a neutralino pair via gluon-gluon
fusion is studied in the minimal supersymmetric model(MSSM) at
proton-proton colliders. The numerical analysis of their production
rates are carried out in the mSUGRA scenario. The results show that
this cross section may reach about 80 femto barn for
 $\tilde{\chi}^{0}_{1}\tilde{\chi}^{0}_{2}$ pair
production and 23 femto barn for $\tilde{\chi}^{0}_{2}\tilde{\chi}^{0}_{2}$
pair production with suitable input parameters at the future LHC collider.
It shows that this loop mediated process can be competitive with the
quark-antiquark annihilation process at the LHC. }\\

\vskip 5mm
{PACS number(s): 14.80.Ly, 12.15.Ji, 12.60.Jv}
\end{minipage}
\end{center}
\end{titlepage}

\baselineskip=0.36in

\eject
\rm
\baselineskip=0.36in

\begin{flushleft} {\bf 1. Introduction} \end{flushleft}
\par
The Standard Model(SM) \cite{sm}\cite{higgs} is a successful
theory of strong and electroweak interactions up to the present accessible
energies. The hierarchy problem suggests that in principle the SM
is the low energy effective theory of a more fundamental one. Supersymmetry
(SUSY) is presently the most popular attempt to solve the hierarchy
problem of the SM, where the cancellation of quadratic divergences is
guaranteed and hence any mass scale is stable under radiative corrections.
The most favorable candidate for a realistic extension of the SM is the
minimal supersymmetric standard model(MSSM). In the MSSM,
a discrete symmetry called R-parity\cite{rparity} is kept in order to
assure baryon and lepton number conservations, because the gauge-coupling
unification supports conservation of R-parity\cite{rcon}. It
implies that there exists an absolutely stable lightest supersymmetric
particle(LSP). Proper electroweak symmetry breaking induces the right
properties of the LSP being a natural weakly interacting cold dark matter
candidate, which can explain many astrophysical observations\cite{darkm}.
In most cases the LSP in the MSSM may be the lightest Majorana fermionic
neutralino. The neutralinos are mass eigenstates which are model dependent
linear combinations of neutral gauginos and higgsinos\cite{haber}. They are
determined by diagonalizing the corresponding mass matrix. In the MSSM
the mass matrix depends on four unknown parameters, namely $\mu$,
$M_2$, $M_1$, and $\tan\beta=v_2/v_1$, the ratio of the vacuum expectation
values of the two Higgs fields. $\mu$ is the supersymmetric Higgs-boson-mass
parameter and $M_2$ and $M_1$ are the gaugino mass parameters associated
with the $SU(2)$ and $U(1)$ subgroups, respectively.
\par
Direct search of supersymmetric particles in experiment is one of the
promising tasks for present and future colliders. The multi-TeV Large
Hadron Collider(LHC) at CERN and the possible future Next Linear Collider(NLC)
are elaborately designed in order to study the symmetry-breaking mechanism
and new physics beyond the SM. When the LEP2 running will be terminated,
the hadron colliders Tevatron and LHC will be the machines left in searching
for supersymmetric particles. Therefore it is necessary to give a proper and
full understanding of the production mechanisms of supersymmetric
particles at hadron colliders. If supersymmetry really exists at TeV scale,
SUSY particles should be discovered and it will be possible to make
accurate measurements to determine their masses and other parameters of the
Lagrangian at the LHC, and then we will have a better understanding of the
supersymmetric model\cite{Han0}. We know that there are several mechanisms
inducing the production of a chargino/neutralino pair at hadron colliders.
One is through the quark-antiquark annihilation called Drell-Yan process,
and another is via gluon-gluon fusion. Although the neutralino pair
production via gluon-gluon fusion is a process induced by one-loop Feynman
diagrams, the production rate can be still significant due to the large
gluon luminosity in hadron colliders. The direct production
channels of chargino/neutralinos and sleptons at the hadron colliders
Tevatron and LHC $p\bar{p}/pp\rightarrow \tilde{\chi}_{i}\tilde{\chi}_{j} +X$
and $\tilde{l}\bar{\tilde{l'}}+X$ via quark-antiquark annihilation are
investigated by Beenakker {\cal et al}\cite{beenakker}. A recent work
\cite{mawg} showed that the one-loop process of the lightest chargino pair
production via gluon-gluon fusion at the LHC can be considered as part of the
NLO QCD correction to the quark-antiquark annihilation process. The neutralino
pair production via quark-antiquark annihilation at the LHC was also
considered by Han {\cal et al.}\cite{han}. The measurement of CP violating
supersymmetric phases in the chargino and neutralino pair production at
the NLC was investigated by Barger {\cal et al}\cite{barger}. The lower
mass limit of $29.1 GeV$ on the lightest neutralino is obtained
experimentally by analyzing the DELPHI results with the assumption that
$M_1/M_2 {}^{>}_{\sim} 0.5$,$tan\beta=1$, $\mu=-62.3 GeV$ and
$M_2=46.0 GeV$\cite{bound}.
\par
In this paper we concentrate on the direct neutralino pair production
via gluon-gluon collisions at the LHC in the framework of the MSSM
with complete one-loop Feynman diagrams. The numerical
calculation will be illustrated in the CP conserving mSUGRA
scenario with five input parameters, namely $m_{1/2}$, $m_0$, $A_0$,
$\mu$ and $\tan\beta$, where $m_{1/2}$, $m_0$ and $A_0$ are the
universal gaugino mass, scalar mass at GUT scale and the trilinear soft
breaking parameter in the superpotential respectively. From these five
parameters, all the masses and couplings of the model are determined by the
evolution from the GUT scale down to the low electroweak scale\cite{msugra}.
The paper is organized as follows: In section 2, we introduce the relevant
features of the model and the analysis of the cross section in this work.
In section 3, we discuss the numerical results of the cross sections and
finally, a short summary is presented.

\begin{flushleft} {\bf 2. The Calculation of $pp\rightarrow gg+X\rightarrow
 \tilde{\chi}^{0}_{i}\tilde{\chi}^{0}_{j}+X$} \end{flushleft}
\par
In the MSSM the physical neutralino mass eigenstates
$\tilde{\chi}^{0}_{i}~~(i=1,2,3,4)$ are the combinations of the neutral
gauginos, $\tilde{B}$ and $\tilde{W}^{3}$, and the neutral higgsino,
$\tilde{H}^{0}_{1}$ and $\tilde{H}^{0}_{2}$. In the two-component fermion
fields $\psi^{0}_{j}= (-i\lambda^{'}, -i\lambda^{3}, \psi_{H^{0}_{1}},
\psi_{H^{0}_{2}})$\cite{haber}, where $\lambda^{'}$ is the bino and
$\lambda^{3}$ is the neutral wino, the neutralino mass term in the
Lagrangian is given by

$$
{\cal L}_{M}=-\frac{1}{2}(\psi^{0})^{T} Y \psi^{0} +h.c.,
\eqno{(2.1)}
$$

where the matrix $Y$ reads

$$
Y=\left (
\begin{array}{cccc}
M_{1} & 0& -m_{Z}\sin\theta_{W}\cos\beta & m_{Z}\sin\theta_{W}\sin\beta\\
0& M_{2} & m_{Z}\cos\theta_{W}\cos\beta & -m_{Z}\cos\theta_{W}\sin\beta\\
-m_{Z}\sin\theta_{W}\cos\beta & m_{Z}\cos\theta_{W}\cos\beta& 0 & -\mu\\
m_{Z}\sin\theta_{W}\sin\beta & -m_{Z}\cos\theta_{W}\sin\beta &-\mu & 0
\end{array}\right)
\eqno{(2.2)}
$$
$M_{1}$, $M_{2}$ and $\mu$ can be complex and introduce CP-violating phases.
By reparametrization of the fields, $M_{2}$ can be real and positive without
loss of generality. The matrix $Y$ is symmetric and can be diagonalized by one
unitary matrix $N$ such that $N_{D}=N^{*}YN^{+}=diag(m_{\tilde{\chi}^{0}_{1}},
m_{\tilde{\chi}^{0}_{2}},m_{\tilde{\chi}^{0}_{3}},m_{\tilde{\chi}^{0}_{4}})$
with the order of $m_{\tilde{\chi}^{0}_{1}}\le m_{\tilde{\chi}^{0}_{2}}\le
m_{\tilde{\chi}^{0}_{3}}\le m_{\tilde{\chi}^{0}_{4}}$. Then the two-component
mass eigenstates can be
$$
\chi^{0}_{i}=N_{ij}\psi^{0}_{j},~~~i, j=1,\dots,4.
\eqno{(2.3)}
$$
The proper four-component mass eigenstates are the neutralinos which are
defined in terms of two-component fields as
$$
\tilde{\chi}^{0}_{i}=\left(\begin{array}{c}
\chi^{0}_{i}\\
\bar{\chi}^{0}_{i}\end{array}\right)
~~~(i=1,\dots, 4),
\eqno{(2.4)}
$$
and the mass term becomes
$$
{\cal L}_{m}=-\frac{1}{2}\sum_{i}\tilde{M}_{i} \bar{\tilde{\chi}^{0}_{i}}
\tilde{\chi}^{0}_{i},
\eqno{(2.5)}
$$
where $\tilde{M}_{i}$ are the diagonal elements of $N_{D}$.
\par
The neutralino pair via gluon-gluon collisions can only be produced through
one-loop diagrams in the lowest order. The calculation for this process can
be simply carried out by summing all unrenormalized reducible and irreducible
diagrams and the result will be finite and gauge invariant. In this work, we
perform the calculation in the 't Hooft-Feynman gauge. The generic Feynman
diagrams contributing to the subprocess
$gg \rightarrow \tilde{\chi}_i^{0} \tilde{\chi}_j^{0}$ in
the MSSM are depicted in Fig.1, where the exchange of incoming
gluons in Fig.1(a.1$~\sim~$3) and Fig.1(c.1$~\sim~$2) are not shown.
All the one-loop diagrams can be divided into three
groups: (1) box diagrams shown in Fig.1(a.1$~\sim~$3), (2) quartic
interaction diagrams in Fig.1(b.1$~\sim~$2), (3) triangle diagrams shown
in Fig.1(c.1$~\sim~$2). The $Z^0$ boson intermediated s-channel diagrams
with quark or squark loops, as shown in Fig.1(b.2), Fig.1(c.1) and Fig.1(c.2),
cannot contribute to the cross section. That can be explained in two fields.
Firstly, the vector component of the $Z^0$ boson wave function does not
couple to the initial gg state as the result of the Laudau-Yang theorem.
Secondly, the CP-odd scalar component of the $Z^0$ boson does not couple
to the invariant CP-even $\tilde{\chi}_i^{0} \tilde{\chi}_j^{0}$ state.
We should mention that there are also some diagrams not
contributing to the process, which we do not draw them in Fig.1(for example
the s-channel diagrams with trilinear gluon interactions). Since the vertices
of $A^{0}(G^{0})-\tilde{q}-\tilde{q}$ vanish\cite{gunion}, there is no
diagrams with a triangle squark loop coupling with an $A^0$ or $G^0$ Higgs
boson.
\par
We denote the reaction of neutralino pair production
via gluon-gluon collision as:
$$
g (p_3, \mu) g (p_4, \nu) \longrightarrow
            \tilde{\chi}_{i}^{0} (p_1) \tilde{\chi}_{j}^{0} (p_2).
 \eqno{(2.6)}
$$
The corresponding matrix element for each of the diagrams
can be written as
$$
{\cal M} = {\cal M}^{\hat{s}}+{\cal M}^{\hat{t}}+{\cal M}^{\hat{u}}
\eqno{(2.7)}
$$
where ${\cal M}^{\hat{s}}$, ${\cal M}^{\hat{t}}$ and ${\cal M}^{\hat{u}}$ are
the matrix elements from s-channel, t-channel and u-channel
diagrams, respectively. The amplitude parts from the u-channel
box and triangle vertex interaction diagrams can be obtained from the
t-channel's by the following:
$$
{\cal M}^{\hat{u}}=(-1)^{\delta_{ij}}{\cal M}^{\hat{t}}(\hat{t}
\rightarrow \hat{u}, p_3 \leftrightarrow p_4, \mu \leftrightarrow \nu),
 \eqno{(2.8)}
$$
where the indices i and j are for the final neutralino. When $i=j$, there is a
relative minus sign due to the Fermi statistics, which requires the amplitude
to be antisymmetric under interchange of the two final identical fermions.
The cross section for this subprocess at one-loop order in unpolarized
gluon collisions can be obtained by
$$
 \hat{\sigma}(\hat{s},gg \rightarrow \tilde{\chi}_{i}^{0}
 \tilde{\chi}_{j}^{0}) = \frac{1}{16 \pi \hat{s}^2}(\frac{1}{2})^{\delta_{ij}}
             \int_{\hat{t}^{-}}^{\hat{t}^{+}} d\hat{t}~
             \sum^{\bf -}_{} |{\cal M}|^2,
\eqno{(2.9)}
$$
where $\hat{t}^\pm=1/2\left[ (m^{2}_{\tilde{\chi}_{i}}+m^{2}_{\tilde{\chi}_{j}}
-\hat{s})\pm \sqrt{(m^{2}_{\tilde{\chi}_{i}}+m^{2}_{\tilde{\chi}_{j}}-\hat{s})^2
-4m^{2}_{\tilde{\chi}_{i}}m^{2}_{\tilde{\chi}_{j}}}\right]$.
The factor $(\frac{1}{2})^{\delta_{ij}}$ is due to the two identical particles
in the final states. The bar over the sum means average over
initial spins and colors.
\par
With the results from Eq.(2.9), we can easily
obtain the total cross section at $pp$ collider by folding the cross
section of the subprocess $\hat{\sigma} (gg \rightarrow
\tilde{\chi}^{0}_i {\tilde{\chi}}^{0}_j)$ with the gluon luminosity,
$$
\sigma(pp \rightarrow gg+X \rightarrow
          \tilde{\chi}^{0}_i \tilde{\chi}^{0}_j+X)=
    \int_{( m_{\tilde{\chi}^{0}_{i}}+m_{\tilde{\chi}^{0}_{j}})^2/s }^{1}
d\tau \frac{d{\cal L}_{gg}}{d\tau} \hat{\sigma}(gg \rightarrow
          \tilde{\chi}^{0}_i \tilde{\chi}^{0}_j,
     \hskip 3mm at \hskip 3mm \hat{s}=\tau s). \eqno{(2.10)}
$$
where $\sqrt{s}$ and $\sqrt{\hat{s}}$  denote the proton-proton
and gluon-gluon c.m.s. energies respectively and
$\frac{d{\cal L}_{gg}} {d\tau}$ is the gluon luminosity, which is defined as
$$
\frac{d{\cal L}_{gg}}{d\tau}=\int_{\tau}^{1}
 \frac{dx_1}{x_1} \left[ f_{g}(x_1,Q^2) f_{g}(\frac{\tau}{x_1},Q^2) \right].
 \eqno{(2.11)}
$$
Here we used $\tau=x_{1}x_{2}$, the definitions of
$x_1$ and $x_2$ are from Ref.\cite{jiang}.
In our numerical calculation we adopt the MRST(mode 2) parton distribution
function $f_{g}(x_{i}, Q^{2})$ \cite{MRST}, and ignore the supersymmetric
QCD corrections to the parton distribution functions for simplicity.
The factorization scale $Q$ was chosen as the average of the final
particle masses $\frac{1}{2}(m_{\tilde{\chi}_{i}}+m_{\tilde{\chi}_{j}})$.
The numerical calculation is carried out for the LHC at the energy $14 TeV$.

\par
\begin{flushleft} {\bf 3. Numerical results and discussions} \end{flushleft}
\par
In this section, we present some numerical results of the total cross
section from the full one-loop diagrams involving virtual (s)quarks for
the process $pp \rightarrow gg +X\rightarrow \tilde{\chi}^{0}_{i}
\tilde{\chi}^{0}_{j}+X$, respectively. The input parameters
are chosen as: $m_t=173.8~GeV$, $m_{Z}=91.187~GeV$, $m_b=4.5~GeV$,
$\sin^2{\theta_{W}}=0.2315$, and $\alpha = 1/128$.
We adopt a simple one-loop formula for the running strong coupling
constant $\alpha_s$
$$
\alpha_s(\mu)=\frac{\alpha_{s}(m_Z)} {1+\frac{33-2 n_f} {6 \pi} \alpha_{s}
              (m_Z) \ln \left( \frac{\mu}{m_Z} \right) }. \eqno(3.1)
$$
where $\alpha_s(m_Z)=0.117$ and $n_f$ is the number of active flavors at
energy scale $\mu$.
In our numerical calculation to get the low energy scenario from the mSUGRA,
the renormalization group equations(RGE's)\cite{RGE} are run from
the weak scale $M_Z$ up to the GUT scale, taking all threshold into account.
We use two loop RGE's only for the gauge couplings and the one-loop
RGE's for the other supersymmetric parameters. The GUT scale boundary
conditions are imposed and the RGE's are run back to $M_Z$, again taking
threshold into account. The decay widths of the intermediate Higgs bosons
are considered at the tree level and these formula can be
found in ref.\cite{hunter}.
\par
The neutralino pair production cross sections dependence on the
factorization scale Q is illustrated in Fig.2, where we choose
the mSUGRA parameters as $m_0=100GeV$, $m_{1/2}=150 GeV$,
$A=300GeV$, $\tan\beta=4$ and $\mu>0$. The masses of $\tilde{\chi}^{0}_{1}$
$\tilde{\chi}^{0}_{2}$ are $51.4GeV$ and $98GeV$. For the
$\tilde{\chi}^{0}_{1}\tilde{\chi}^{0}_{2}$ pair production, the cross section
changes from $28 fb$ to $30fb$ when the scale Q changes from
$0.2 m_{\tilde {x}^{0}_{2}}$ to $m_{\tilde {x}^{0}_{2}}$. For the
$\tilde{\chi}^{0}_{2} \tilde{\chi}^{0}_{2}$ pair production, the cross
section changes from $14 fb$ to $15.5 fb$ when the Q changes from
$0.2 m_{\tilde {x}^{0}_{2}}$ to $m_{\tilde {x}^{0}_{2}}$. When Q is larger
than $m_{\tilde {x}^{0}_{2}}$, the cross sections of these two processes
are nearly independent of the factorization scale.
\par
In Fig.3 we show the cross sections of gaugino-like neutralino
pair production and the higgsino-like neutralino pair production. For the
gaugino-like neutralino pair production, we choose the same values of
mSUGRA parameters adopted in Fig.2. But in the higgsino-like neutralino
case, we choose $M_1=150GeV$, $M_2=210 GeV$ and $\mu=90GeV$ in
order to keep the masses of $\tilde{\chi}^{0}_{1}$ and $\tilde{\chi}^{0}_{2}$
being the same as those in the previous gaugino-like neutralino case, the other
parameters of the higgsino-like neutralino case at the weak scale are chosen
being the same as those in the gaugino-like neutralino case. Then the
difference of the cross sections between gaugino-like neutralino pair
production and higgsino-like neutralino pair production only comes from
the change of the coupling of $\tilde{\chi}^{0}_{i}-\tilde{q}-q$ and
$\tilde{\chi}^{0}_{i}- \tilde{\chi}^{0}_{j}-H^{0}(A^0, h^0, G^0)$ where
the matrix elements of N are changed. For the $\tilde{\chi}^{0}_{1}
\tilde{\chi}^{0}_{2}$ pair production, the cross section in the
higgsino-like case is smaller than that in the gaugino-like case,
but for the $\tilde{\chi}^{0}_{2}\tilde{\chi}^{0}_{2}$ pair production,
the cross section in the higgsino-like case is larger than that in the
gaugino-like case.
\par
The cross sections for $\tilde{\chi}^{0}_{1}\tilde{\chi}^{0}_{2}$
and $\tilde{\chi}^{0}_{2}\tilde{\chi}^{0}_{2}$ productions at the LHC versus
the mass of $\tilde{\chi}^{0}_{2}$ is shown in Fig.4. We assume that
$\tilde{\chi}^{0}_{1}$ is the LSP and escapes detection.
The input parameters are chosen as $m_{0}=100GeV$,
$A_{0}= 300GeV$, $\tan\beta=4$, $\mu>0$ and $m_{1/2}$ varying from $130GeV$
to $325GeV$. In the framework of the mSUGRA, the masses of
$\tilde{\chi}^{0}_{1}$ and $\tilde{\chi}^{0}_{2}$ increase from
$41GeV$ to $129GeV$ and from $81GeV$ to $251GeV$ respectively, and the
masses of Higgs boson $H^0$ and $A^0$ increase from $230GeV$ to $509GeV$
and from $226GeV$ to $507GeV$ respectively with our input parameters.
In our case, the mass of the Higgs boson $h^0$ is always smaller than
$m_{\tilde{\chi}^{0}_{1}} +m_{\tilde{\chi}^{0}_{2}}$ and the contribution to
the cross section of $h^0$ is very small due to the s-channel suppression.
Since the masses of $H^0$ and $A^0$ are larger than $m_{\tilde{\chi}^{0}_{1}}
+ m_{\tilde{\chi}^{0}_{2}}$, the cross section will be strongly enhanced
by the s-channel resonance effects when the total energy $\sqrt{\hat{s}}$ of
the subprocess approaches the mass of $H^0$ or $A^0$. Since
$ m_{\tilde{\chi}^{0}_{2}} >m_{\tilde{\chi}^{0}_{1}}$, the pair production
of $\tilde{\chi}^{0}_{2}\tilde{\chi}^{0}_{2}$ compared to $\tilde{\chi}^{0}
_{1} \tilde{\chi}^{0}_{2}$ is kinematically suppressed. For the
$\tilde{\chi}^{0}_{1}\tilde{\chi}^{0}_{2}$ pair production, the cross section
can reach 43 femto barn when $m_{1/2}=130GeV$,
$m_{\tilde{\chi}^{0}_{1}} =41GeV$ and $m_{\tilde{\chi}^{0}_{2}}=81GeV$. The
cross section of $\tilde{\chi}^{0}_{2}\tilde{\chi}^{0}_{2}$ pair production
can reach 23 femto barn when $m_{1/2}=130~GeV$. From the results of
\cite{beenakker}, we know that the cross sections of
$\tilde{\chi}^{0}_{1}\tilde{\chi}^{0}_{2}$ and
$\tilde{\chi}^{0}_{2}\tilde{\chi}^{0}_{2}$ pair production via
quark-antiquark annihilation at the LHC are about 70 femto barn and 300 femto
barn respectively, when the input parameters are chosen as $m_{0}=150~GeV$,
$A_{0}= 300~GeV$, $\tan\beta=4$, $\mu>0$ and $m_{\tilde{\chi}^{0}_{2}}=81~GeV$.
The pair production rate of $\tilde{\chi}^{0}_{1}\tilde{\chi}^{0}_{2}$ via
gluon-gluon fusion is about few ten percent of that via
quark-antiquark annihilation at the LHC with the same input parameters. The
cross section of $\tilde{\chi}^{0}_{2} \tilde{\chi}^{0}_{2}$  via gluon-gluon
fusion is about few percent of that via quark-antiquark annihilation at the
LHC. The neutralino pair production via gluon-gluon fusion at the LHC is of
the same order of the next leading order(NLO) QCD correction to the
quark-antiquark production process. The neutralino pair production rate via
gluon-gluon fusion should be considered in studying the pair pair productions
of neutralinos at proton-proton colliders.
\par
In Fig.5 we present the cross sections of neutralino pair productions versus
$\tan\beta$ where the other four input parameters are chosen as $m_{0}=100~GeV$,
$A_{0}= 300~GeV$, $\mu>0$ and $m_{1/2}=150~GeV$. When $\tan\beta$ increases
from 4 to 32, the mass of $\tilde{\chi}^{0}_{1}$ increases from $51~GeV$
to $57~GeV$ and the mass of $\tilde{\chi}^{0}_{2}$ increases from $98~GeV$
to $102.5~GeV$. So the masses of neutralinos are nearly independent of
$\tan\beta$. The masses of Higgs boson $H^0$ and $A^0$ depend on
$\tan\beta$ and decrease from $258~GeV$ to $165~GeV$ and from $254~GeV$ to
$165~GeV$ respectively. The mass of $h^0$ is a function of $\tan\beta$ too,
but keeps $m_{h^0}<m_{\tilde{\chi}^{0}_{1}} + m_{\tilde{\chi}^{0}_{2}}$.
Since the couplings of Higgs bosons to quarks and squarks pair are related
to the ratio of the vacuum expectation values, $\tan\beta$ should effect
the cross sections substantially. The cross sections of the neutralino pair
productions decrease when $\tan\beta$ goes from 4 to 6, and increase when
$20>\tan\beta>6$ due to the effect of the coupling strength of Higgs bosons
to quarks and squarks pair. For the $\tilde{\chi}^{0}_{1}\tilde{\chi}^{0}_{2}$
pair production, when $tan\beta<27$, the masses of $H^0$ and $A^0$ are larger
than $m_{\tilde{\chi}^{0}_{1}} + m_{\tilde{\chi}^{0}_{2}}$, the s-channel
resonance effects of $H^0$ and $A^0$ on the cross section can be enhanced,
while when $tan\beta>27$, the masses of $H^0$ and $A^0$ become smaller than
$m_{\tilde{\chi}^{0}_{1}} + m_{\tilde{\chi}^{0}_{2}}$, then there will be no
s-channel resonance effects and the cross section goes down rapidly. For the
similar reason, the cross section of the $\tilde{\chi}^{0}_{2}
\tilde{\chi}^{0}_{2}$ pair production increase in the range of
$20>\tan\beta>6$ due to the effect of the Yukawa couplings of Higgs bosons
to quarks and squarks pair and the s-channel resonance effects of the
intermediate Higgs bosons, but it goes down again when $\tan\beta>20$ where
the masses of $H^0$ and $A^0$ become smaller than $m_{\tilde{\chi}^{0}_{2}}
+ m_{\tilde{\chi}^{0}_{2}}$ and the s-channel resonance effects of the
Higgs bosons disappear. From the figure we can see that if we choose
suitable input parameters, the cross section can be largely enhanced.
Especially the cross section of the $\tilde{\chi}^{0}_{1}
\tilde{\chi}^{0}_{2}$ pair production can even reach about 80 femto barn
when $\tan\beta \sim 27$.
\par
In our calculation we find that the discrepancy between our results from the
newer sets of gluon densities in the MRST(mode 2) scheme and the older
sets in the MRS(set G) scheme\cite{MRS} is very small and cannot be
distinguished in figures.

\par
\begin{flushleft} {\bf 4. Summary} \end{flushleft}
\par
In this paper, we studied the pair production process of the neutralino
via gluon-gluon fusion at the LHC collider. The numerical
analysis of their production rates is carried out in the mSUGRA scenario
with some typical parameter sets. The results show that the cross section
of the neutralino pair-production via gluon-gluon fusion can reach
about 80 femto barn for $\tilde{\chi}^{0}_{1}\tilde{\chi}^{0}_{2}$ pair
production and 23 femto barn for $\tilde{\chi}^{0}_{2}\tilde{\chi}^{0}_{2}$
pair production at the LHC collider. In some c.m.s energy
regions of incoming gluons, where the s-channel resonance conditions are
satisfied in the parameter space, we can see observable enhancement effects
of the cross sections. We conclude that neutralino pair production via
gluon-gluon fusion can be competitive with the quark-antiquark annihilation
production process at the LHC and should be considered as a part of the NLO QCD
correction to the quark-antiquark annihilation production process.

\vskip 4mm
\noindent{\large\bf Acknowledgement:}
This work was supported in part by the National Natural Science
Foundation of China(project number: 19875049), the Youth Science
Foundation of the University of Science and Technology of China, a grant
from the Education Ministry of China and the exchange program between
China and Austria(Project number V.A.12). One of the authors Y. Jiang
would like to thank Prof. F.F. Sch\"{o}berl for valuable discussion in this
context.

\vskip 10mm
\begin{flushleft} {\bf Figure Captions} \end{flushleft}

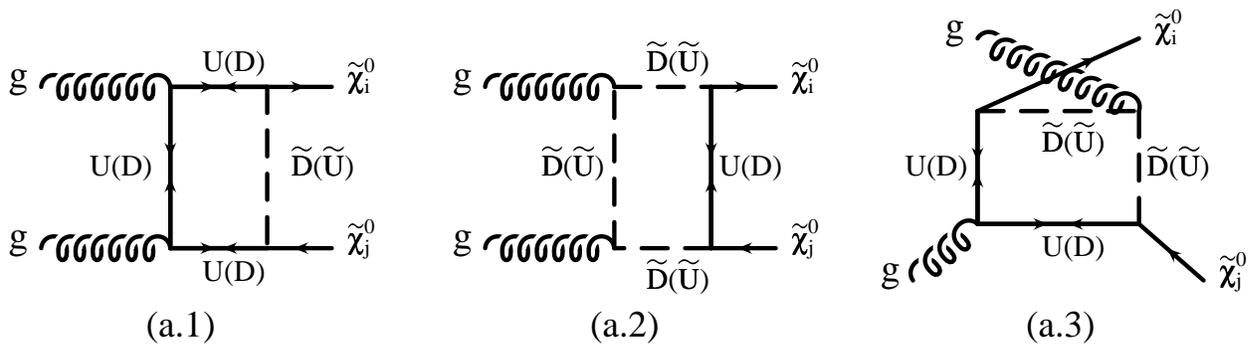
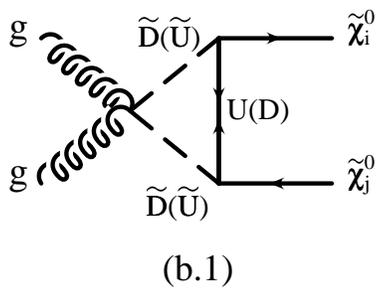
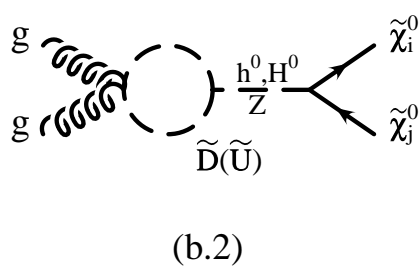
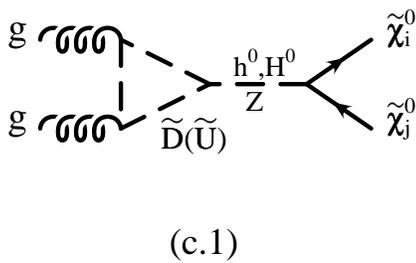
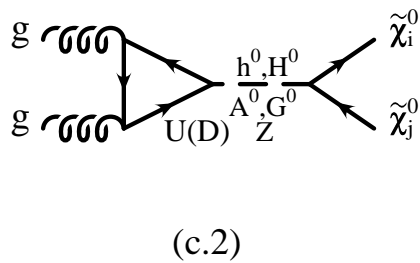
{\bf Fig.1} The Feynman diagrams of the subprocess $g g \rightarrow
\tilde{\chi}^{0}_{i}\tilde{\chi}^{0}_{j}$.

{\bf Fig.2} Dependence of the cross sections for the productions of
$\tilde{\chi}^{0}_{1}\tilde{\chi}^{0}_{2}$ and
$\tilde{\chi}^{0}_{2}\tilde{\chi}^{0}_{2}$ pairs on the factorization
scale Q with the mSUGRA parameters $m_0=100~GeV$, $m_{1/2}=150~GeV$,
$A=300~GeV$, $\tan\beta=4$ and $\mu>0$. The masses of $\tilde{\chi}^{0}_{1}$
$\tilde{\chi}^{0}_{2}$ are $51.4~GeV$ and $98~GeV$.

{\bf Fig.3} The cross sections for the neutralino pair production at the
gaugino-like neutralino case and higgsino-like neutralino case at the LHC.
For the gaugino-like neutralino case , the parameters are chosen in mSUGRA
scenario with $m_0=100~GeV$, $m_{1/2}=150~GeV$, $A=300~GeV$, $\tan\beta=4$
and $\mu>0$. For the higgsino-like neutralino case, we choose $M_1=150~GeV$,
$M_2=210~GeV$, $\mu=90~GeV$ and the other parameters the same as those in
the gaugino-like neutralino case.

{\bf Fig.4} Total cross sections of the process $p p\rightarrow
g g+X \rightarrow \tilde{\chi}^{0}_{i}\tilde{\chi}^{0}_{j}+X$ as function of
$m_{\tilde{\chi}^{0}_{2}}$ at the LHC. The solid curve is for
$\tilde{\chi}^{0}_{1}\tilde{\chi}^{0}_{2}$ pair
production and the dashed curve is for
$\tilde{\chi}^{0}_{2}\tilde{\chi}^{0}_{2}$ pair production.

{\bf Fig.5} Total cross sections of the process $p p\rightarrow
g g +X \rightarrow \tilde{\chi}^{0}_{i}\tilde{\chi}^{0}_{j}+X$ as function of
$\tan\beta$ at the LHC.
The solid curve is for $\tilde{\chi}^{0}_{1}\tilde{\chi}^{0}_{2}$ pair
production and the dashed curve is for
$\tilde{\chi}^{0}_{2}\tilde{\chi}^{0}_{2}$ pair production.

\end{large}
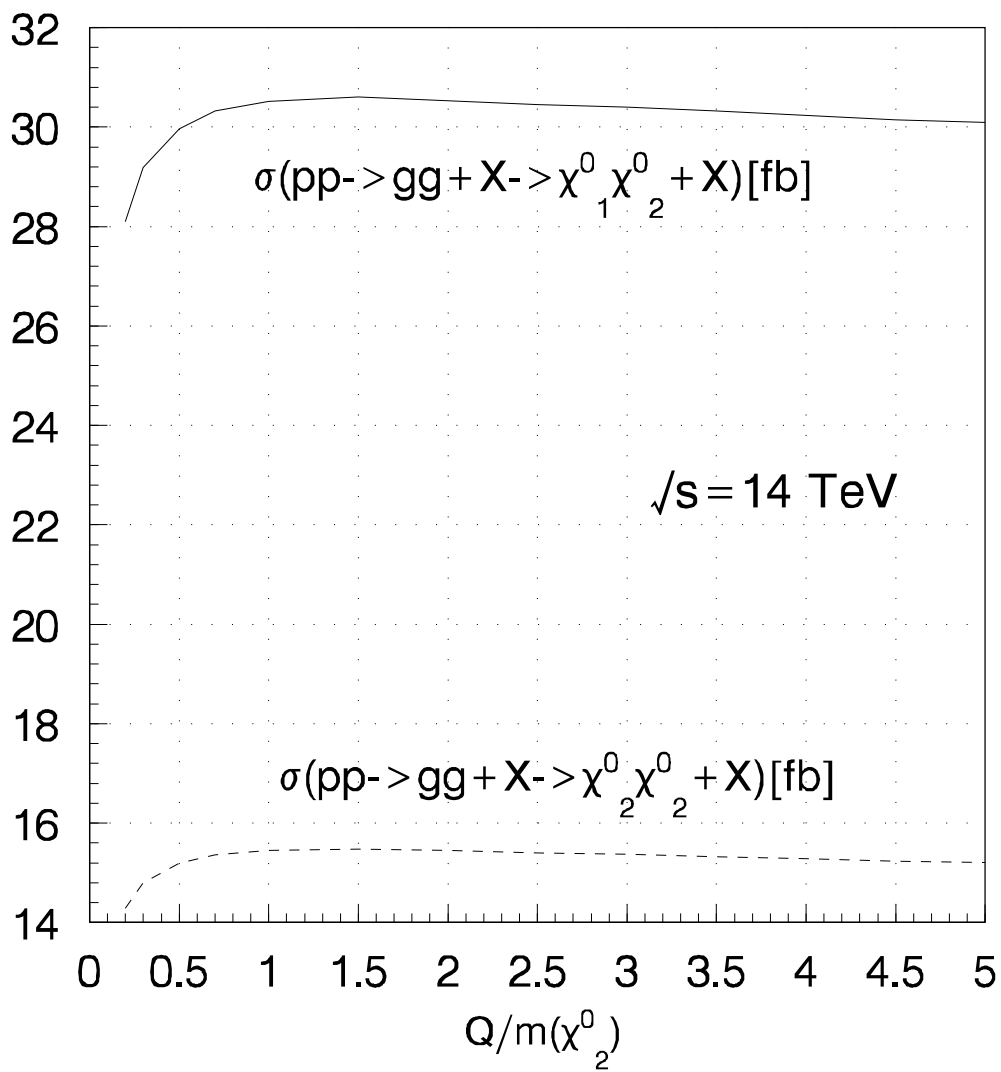
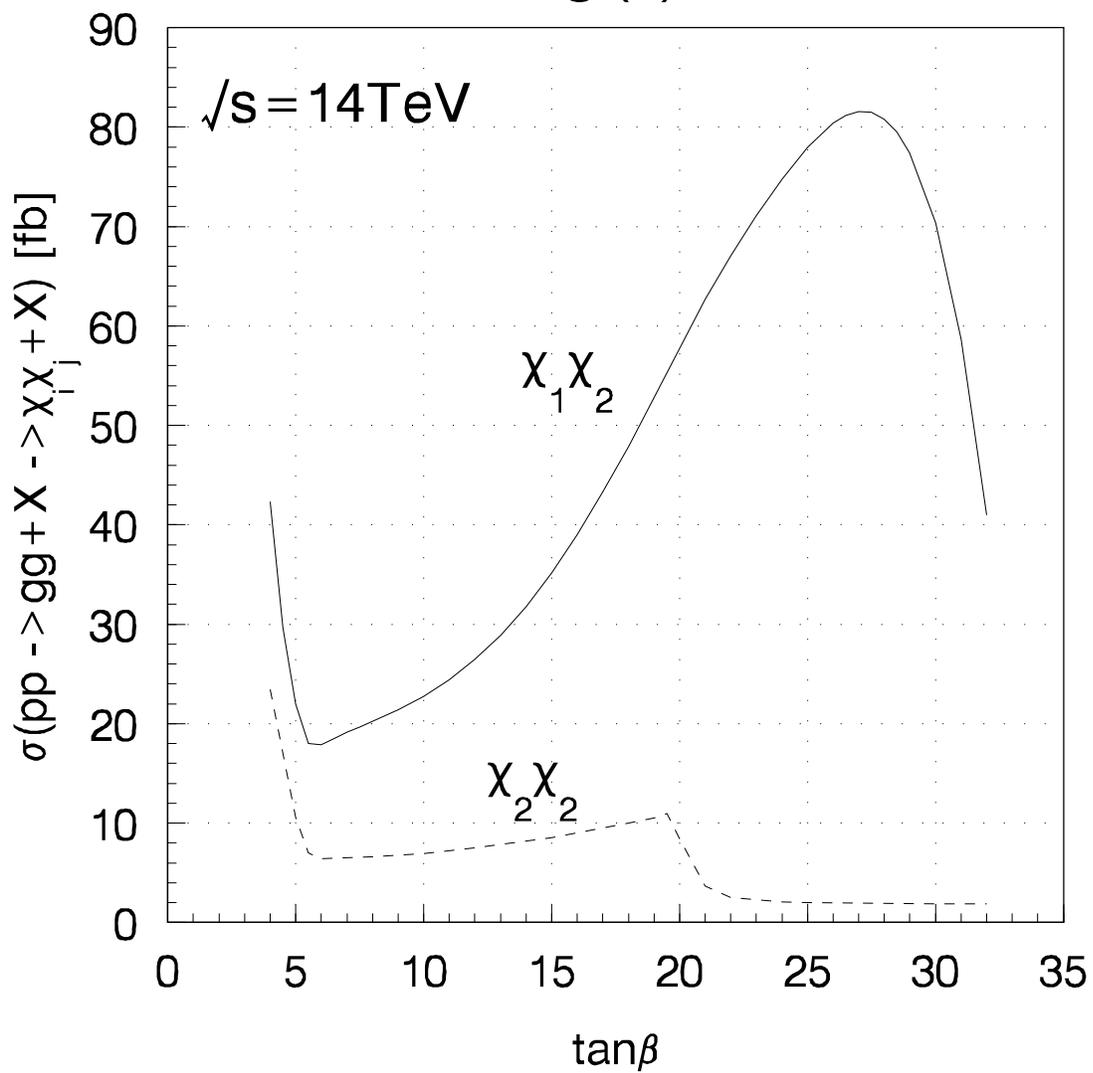
\end{document}